# Relativistic Ginzburg–Landau equation: An investigation for overdoped cuprate films

Yong Tao[†]

College of Economics and Management, Southwest University, Chongqing, China

Department of Management, Technology and Economics, ETH Zurich, Switzerland

**Abstract:** By introducing the imaginary time, Gor'kov's Ginzburg–Landau equation at zero temperature can be extended to an exact relativistic form without any phenomenological parameter, which is intended to describe the zero-temperature overdoped cuprate. By using such a relativistic equation, we have shown that the two-class scaling observed in the overdoped side of single-crystal $La_{2-x}Sr_xCuO_4$ (LSCO) films [Nature **536**, 309–311 (2016)] can be derived exactly. In this paper, we further test the validity of the relativistic Ginzburg–Landau equation. By applying the perturbation method into this equation, we theoretically predict that near the superconductor–metal transition point in the overdoped side of LSCO films, the zero-temperature correlation length $\xi(0)$ and the transition temperature $T_c$ should yield a novel scaling $\xi(0) \propto T_c^{-\sigma}$ with a critical exponent $\sigma \approx 1.31$ (up to the two-loop approximation). Here, we propose a diffraction experiment between $X$-rays and zero-temperature LSCO films to measure the critical exponent $\sigma$.

---

[†] Correspondence to: taoyingyong@yahoo.com

# 1. Introduction

Quantum critical phenomena (QCP) are a central topic in condensed matter physics. Different from thermal critical phenomena, QCP are a new kind of phase transitions that are driven, not by thermal fluctuations, but by quantum fluctuations associated with Heisenberg's uncertainty principle. This means that QCP should occur around (absolute) zero temperature, where the thermal fluctuations can be ignored. The theoretical framework for understanding the QCP is due to Hertz's seminal work [1]. In Hertz's theory, quantum mechanics appears by introducing an imaginary time to describe the dynamics of strongly correlated systems near the zero temperature. Later, Hertz's theory was extended by Mills [2]. The zero-temperature superfluid phase stiffness $\rho_s(0)$ and the transition temperature $T_c$ are two important parameters for investigating the QCP in superconductors, It has been found that the superconductor–insulator transition in the highly underdoped side of the cuprate might be a kind of QCP, where $T_c$ and $\rho_s(0)$ yield a sub-linear relationship [3–6]. However, there is scant understanding of the superconductor–metal transition occurring in the highly overdoped side of the cuprate.

Recently, by investigating the overdoped side of single-crystal $La_{2-x}Sr_xCuO_4$ (LSCO) films, Bozovic *et al.* observed that $T_c$ and $\rho_s(0)$ obeyed two-class scaling [7]:

$$\begin{cases} T_c = \alpha \cdot \rho_s(0) + T_0, & T_c \geq T_M \\ T_c = \gamma \cdot \sqrt{\rho_s(0)}, & T_c \leq T_Q \end{cases} \quad (1)$$

where $T_M \approx 12$ K, $T_Q \approx 15$ K, $\alpha = 0.37 \pm 0.02$, $T_0 = (7.0 \pm 0.1)$ K, and $\gamma = (4.2 \pm 0.5)$ K$^{1/2}$.

The linear part in equation (1) has been well known as Homes' law [8–9], which is a result of the mean-field theory of Abrikosov–Gor'kov [10]. As evidence, by using this mean-field theory, Khodel et al. [11] have produced the correct theoretical value of $\alpha$. Nevertheless, equation (1) indicates that a parabolic scaling emerges in the highly overdoped side. Such a parabolic scaling differs significantly from Homes' law; therefore, Bozovic *et al.* concluded that their experimental findings are incompatible with the mean-field description [7, 12–13]. As possible evidence, Bozovic *et al.* have observed that, with increased doping ($T_c \to 0$), LSCO becomes more metallic, and there is induction of a quantum phase transition from a superconductor to a normal metal [12–13]. To understand the parabolic scaling in equation (1), by introducing the imaginary time [1, 14], we [15–17] extended Gor'kov's Ginzburg–Landau equation at zero temperature to an exact relativistic form, which is intended to describe the zero-temperature overdoped cuprate. Our calculation shows that [15–17], as $T_c \to 0$, quantum fluctuations play an important role in inducing the parabolic scaling, and that the two-class scaling (1) can be derived exactly. As evidence justifying the relativistic Ginzburg–Landau equation, we have obtained the correct theoretical values of $\gamma$, $T_M$, and $T_Q$ [16–17]. In this paper, we further test the validity of the relativistic Ginzburg–Landau equation, and we theoretically predict that near the superconductor–metal transition point in the overdoped side of LSCO films, the zero-temperature correlation length $\xi(0)$ and the transition temperature $T_c$ should yield a novel scaling $\xi(0) \propto T_c^{-\sigma}$ with a critical exponent $\sigma \approx 1.31$ (up to the two-loop approximation). Finally, we propose an experiment for measuring the critical exponent $\sigma$. Here, we adopt the natural units $\hbar = c = k_B = 1$, where $\hbar$ denotes the reduced Planck constant, $c$ is the light speed, and $k_B$ is the Boltzmann constant.

## 2. Relativistic Ginzburg–Landau equation

Let us first introduce the relativistic form of the Ginzburg–Landau equation in the imaginary time formalism [16–17]. By using the BCS Hamiltonian of superconductivity, Gor'kov has shown that when $|T - T_c| \approx 0$, the Ginzburg–Landau equation can be written in the form [10]:

$$\frac{1}{4m_e^*}\nabla^2\psi(T) - \frac{1}{\lambda} \cdot \frac{(T-T_c)}{T_c}\psi(T) - \frac{1}{\lambda \cdot n_s(0)}|\psi(T)|^2\psi(T) = 0, \qquad (2)$$

where $\lambda = \frac{7\zeta(3)\cdot\varepsilon_F}{6\pi^2 T_c^2}$ and $|\psi(T)|^2$ denote the superfluid density at the temperature $T$. Moreover, $n_s(0)$ denotes the zero-temperature superfluid density when materials are homogenous, $\zeta(x)$ is the Riemann zeta function, $\varepsilon_F$ is the Fermi energy, and $m_e^*$ is the mass of an electron. By rescaling $\psi(T)$ according to $\phi(T) = \frac{1}{\sqrt{4m_e^*}}\psi(T)$, equation (2) yields the following Lagrangian function:

$$\mathcal{L}(T) = |\boldsymbol{\nabla}\phi(T)|^2 + \frac{24\pi^2 m_e}{7\zeta(3)\cdot\varepsilon_F}T_c^2 \cdot \frac{(T-T_c)}{T_c}\cdot|\phi(T)|^2 + \frac{12\pi^2 m_e}{7\zeta(3)\cdot\varepsilon_F}\cdot\frac{T_c^2}{\rho_s(0)}\cdot|\phi(T)|^4, \qquad (3)$$

where $\rho_s(0) = \frac{n_s(0)}{4m_e^*}$.

Equation (3) can be used to describe thermal critical phenomena in the case of finite temperature ($T > 0$). In this case, the coefficient in front of $|\phi(T)|^2$ is linear in $(T - T_c)$. The partition function of equation (3) is written as below [18–19]:

$$Z(T) = \int[\mathcal{D}\phi^*(T)]_\Lambda \int[\mathcal{D}\phi(T)]_\Lambda \, e^{-\frac{1}{T}\int d^D q \cdot \mathcal{L}(T)}, \qquad (4)$$

where $\Lambda = 1/a$ denotes the momentum cut-off, with $a$ being the lattice spacing, and $\boldsymbol{q}$ denotes the spatial coordinates.

Over recent decades, with the great advances in cooling technologies, much attention has been focused on investigating the superconducting materials that are extremely close to $T = 0$. For example, regarding overdoped LSCO films, Bozovic *et al.* had reduced $T$ to a level below 0.3 K and kept $T_c > 5$ K [7]. In this regard, compared to $T_c$, $T$ can be taken as approximately 0 K. This kind of superconducting material can, therefore, be approximately described by $\mathcal{L}(T = 0)$, which, by equation (3), yields:

$$\mathcal{L}(T = 0) = |\boldsymbol{\nabla}\phi(0)|^2 - \frac{24\pi^2 m_e}{7\zeta(3)\cdot\varepsilon_F}T_c^2 \cdot |\phi(0)|^2 + \frac{12\pi^2 m_e}{7\zeta(3)\cdot\varepsilon_F}\cdot\frac{T_c^2}{\rho_s(0)}\cdot|\phi(0)|^4. \qquad (5)$$

Different from the case of $T > 0$, the coefficient in front of $|\phi(0)|^2$ is quadratic in $T_c$, rather than being linear in $(T - T_c)$. Furthermore, the coefficient in front of $|\phi(0)|^4$ is also quadratic in $T_c$. More importantly, $T = 0$ is a singular point in equation (4); therefore, one cannot deal with the case of zero temperature by simply substituting equation (5) into equation (4). To take the zero-temperature case into account, we need to introduce an imaginary time $\tau \in [0, 1/T]$ with $T = 0$ [1, 14]. In the imaginary time formalism, the order parameter $\phi(0)$ is a function of space and imaginary time [1, 14, 15–17]; that is, $\phi(0) = \phi(\boldsymbol{q}, \tau)$. Therefore, we should add the possible time-derivative terms [14] $\phi^*(\boldsymbol{q},\tau)\partial_\tau\phi(\boldsymbol{q},\tau)$ or $|\partial_\tau\phi(\boldsymbol{q},\tau)|^2$ into the equation (5). Here, we propose to add the quadratic term of time-derivative

$|\partial_\tau \phi(\boldsymbol{q},\tau)|^2$ so that equation (5) yields an exact relativistic form:

$$\mathcal{L}_q(T=0) = |\partial_\tau \phi(\boldsymbol{q},\tau)|^2 + |\boldsymbol{\nabla}\phi(\boldsymbol{q},\tau)|^2 - \frac{24\pi^2 m_e}{7\zeta(3)\cdot\varepsilon_F} T_c^2 \cdot |\phi(\boldsymbol{q},\tau)|^2 + \frac{12\pi^2 m_e}{7\zeta(3)\cdot\varepsilon_F} \cdot \frac{T_c^2}{\rho_s(0)} \cdot |\phi(\boldsymbol{q},\tau)|^4, \quad (6)$$

where the coefficient in front of $|\partial_\tau \phi(\boldsymbol{q},\tau)|^2$ has been set to be 1. Based on this setting, there are no phenomenological parameters in the relativistic equation (6), by which one can obtain the relativistic Ginzburg–Landau equation in the imaginary time formalism [16–17]. We will immediately justify the validity of the relativistic ansatz $|\partial_\tau \phi(\boldsymbol{q},\tau)|^2$ in equation (6).

Substituting equation (6) into equation (4) we obtain the quantum partition function:

$$Z(T=0) = \int [\mathcal{D}\phi^*(\boldsymbol{q},\tau)]_\Lambda \int [\mathcal{D}\phi(\boldsymbol{q},\tau)]_\Lambda \, e^{-\int d\tau \int d^D q \cdot \mathcal{L}_q(T=0)}, \quad (7)$$

which is intended to describe the zero-temperature overdoped cuprate [16–17].

To justify the validity of the relativistic ansatz $|\partial_\tau \phi(\boldsymbol{q},\tau)|^2$ in equation (6), we first show that by using equation (7) one can reproduce the experimental result (1). To this end, let us denote the coefficients in front of $|\phi(\boldsymbol{q},\tau)|^2$ and $|\phi(\boldsymbol{q},\tau)|^4$ by $\lambda_2(T_c)$ and $\lambda_4(T_c)$, respectively. Therefore, by equation (6) we have:

$$\lambda_2(T_c) = -\frac{24\pi^2 m_e}{7\zeta(3)\cdot\varepsilon_F} T_c^2, \quad (8)$$

$$\lambda_4(T_c) = \frac{12\pi^2 m_e}{7\zeta(3)\cdot\varepsilon_F} \cdot \frac{T_c^2}{\rho_s(0)}. \quad (9)$$

For overdoped LSCO films at $T=0$, we observe that $T_c = 0$ is a quantum critical point [7]. Therefore, as $T_c \to 0$, quantum fluctuations are expected to be amplified so that the mean-field approximation breaks down. Based on this observation, by applying the renormalization group approach into the quantum partition function (7), we have proved that [15–17] when $T_c \to 0$, $\lambda_4(T_c)$ yields a fixed point depending on the momentum cut-off $\Lambda$ (up to the one-loop correction):

$$\lambda_4(T_c) = \frac{12\pi^2 m_e}{7\zeta(3)\cdot\varepsilon_F} \cdot \frac{T_c^2}{\rho_s(0)} = (4-D-z) \cdot \Lambda^{4-D-z} \cdot \frac{(2\pi)^D \Gamma\left(\frac{D}{2}\right)}{5(\pi)^{\frac{D}{2}}}, \quad (10)$$

where $z$ is the quantum dynamical exponent. The relativistic ansatz $|\partial_\tau \phi(\boldsymbol{q},\tau)|^2$ in equation (6) indicates $z=1$. Therefore, for $D=2$ (LSCO films), equation (10) leads to a parabolic scaling [15–17]:

$$T_c = \gamma(2) \cdot \sqrt{\rho_s(0)}, \tag{11}$$

where $\gamma(2) = \sqrt{\frac{7 \cdot \zeta(3) \cdot \varepsilon_F}{15 \cdot \pi \cdot a \cdot m_e}}$.

By contrast, the non-relativistic term $\phi^*(\mathbf{q},\tau)\partial_\tau \phi(\mathbf{q},\tau)$ indicates $z = 2$, which by equation (10) implies that $\lambda_4(T_c) = 0$ for $D = 2$; that is, the parabolic scaling (11) vanishes. Therefore, the relativistic ansatz $|\partial_\tau \phi(\mathbf{q},\tau)|^2$ in equation (6) is a prerequisite for obtaining the parabolic scaling (11). This justifies the relativistic ansatz $|\partial_\tau \phi(\mathbf{q},\tau)|^2$ in equation (6), as, by equation (1), the parabolic scaling (11) has been experimentally observed. Next, we further point out that $\gamma(2)$ agrees with the experimental measure value.

We have clarified that [15–17] the parabolic scaling (11) holds if and only if the mean-field approximation breaks down. To identify the scope of application of the mean-field approximation, we have proposed a quantum Ginzburg number to derive the following result [17]:

$$\begin{cases} T_c = \alpha \cdot \rho_s(0) + T_0, & T_c \geq T_M \approx \frac{T_0}{1-\alpha} \\ T_c = \gamma(2) \cdot \sqrt{\rho_s(0)}, & T_c \leq T_Q \approx \gamma(2)^2 \end{cases}. \tag{12}$$

Equation (12) indicates that the parabolic scaling (11) holds when $T_c \leq \gamma(2)^2$. In particular, when $T_c \geq T_0/(1-\alpha)$, the mean-field approximation is valid, and then the relationship between $T_c$ and $\rho_s(0)$ conforms to Homes' linear law [8–10, 15–17]. Regarding overdoped LSCO films, substituting the data $a \approx 3.8 \times 10^{-10}$ m, $\alpha \approx 0.37$, $T_0 \approx 7$ K [7], and $\varepsilon_F(x \approx 0.2) \approx 8.75$ eV [20] into equation (12), we obtain the theoretical values:

$$\gamma(2) \approx 4.29 \text{ K}^{1/2}, \tag{13}$$

$$T_M \approx 11 \text{ K}, \tag{14}$$

$$T_Q \approx 17 \text{ K}, \tag{15}$$

which are well in accordance with the experimental measure values in equation (1). In particular, the theoretical result (13) holds when the coefficient in front of the quadratic term $|\partial_\tau \phi(\mathbf{q},\tau)|^2$ in equation (6) is equal to 1. This is strong evidence supporting the validity of the relativistic equation (6). In this paper, we further show that the relativistic equation (6) leads to a new theoretical prediction that can be tested.

## 3. Theoretical result

In general, quantum critical phenomena are produced by $T$-driven transitions or $x$-driven transitions [21], where $x$ denotes the tuning parameter. In this paper, we investigate the zero-temperature $x$-driven transitions with $x$ being the doping level. If one denotes the overdoped limit by $x_o$, one should expect a scaling [22–23]:

$$T_c \propto (x_o - x)^m. \tag{16}$$

For overdoped LSCO, the existing experimental investigation showed [22] $m = 1/2$; refer to Figure 1.

**[Insert Figure 1 here]**

Equation (16) indicates that one can control the parameter $x$ (doping level) to drive the transition temperature $T_c$ to zero, and it constitutes a singular limit for the ensuing quantum phase transition. Substituting equation (16) into equation (8) yields:

$$\lambda_2(T_c) \propto T_c^2 \propto (x_o - x)^{2m}. \tag{17}$$

Different from the $T$-driven transitions, where $\lambda_2$ is linear in $(T - T_c)$, equation (17) indicates that $\lambda_2(T_c)$ is quadratic in $T_c$. Now, due to this difference, we show that equation (17) leads to a novel scaling relationship between the transition temperature $T_c$ and the zero-temperature correlation length $\xi(0)$. To this end, we assume that quantum fluctuations with wavelengths larger than $2\pi/\Lambda$ cannot be averaged out. By the theory of critical phenomena, this means that the coefficients $\lambda_2(T_c)$ and $\lambda_4(T_c)$ in equation (6) should receive the contributions from quantum fluctuations at these size scales. Thus, by applying the renormalization group method into the quantum partition function (7), we obtain the following equations [15]:

$$\lambda_2'(b) = b^2 \left[ \lambda_2 + 4\lambda_4 \int_{-\infty}^{\infty} \frac{d\omega}{2\pi} \int_{\Lambda/b}^{\Lambda} \frac{d^D k}{(2\pi)^D} \left( \frac{1}{\omega^2 + k^2 + \lambda_2} - \frac{1}{\omega^2 + k^2} \right) \right], \tag{18}$$

$$\lambda_4'(b) = b^{4-D-z} \lambda_4 \left[ 1 - 10\lambda_4 \int_{-\infty}^{\infty} \frac{d\omega}{2\pi} \int_{\Lambda/b}^{\Lambda} \frac{d^D k}{(2\pi)^D} \left( \frac{1}{\omega^2 + k^2 + \lambda_2} \right)^2 \right], \tag{19}$$

where $b$ denotes the parameter that guarantees the rescaling transformation $\mathbf{q}' = b^{-1}\mathbf{q}$, and $\tau' = b^{-z}\tau$. Due to the relativistic ansatz $|\partial_\tau \phi(\mathbf{q}, \tau)|^2$ in equation (6) we have $z = 1$ [15]. By using equations (18) and (19), we can obtain:

$$\lambda'_2(b) \approx \lambda_2 b^{2-4\hat{\lambda}_4}, \tag{20}$$

$$\hat{\lambda}'_4(b) \approx \hat{\lambda}_4 b^{4-D-z-10\hat{\lambda}_4}, \tag{21}$$

where $\hat{\lambda}_4 = \lambda_4 \frac{S_D \Lambda^{D-3}}{4(2\pi)^D}$ and $S_D = 2\pi^{D/2}/\Gamma(D/2)$.

The derivation for equations (20) and (21) can be found in the Appendix. Equation (20) indicates $b \approx (\lambda'_2(b)/\lambda_2)^{1/(2-4\hat{\lambda}_4)}$. For $b \gg 1$, we choose the parameter $b$ so that $\lambda'_2(b) = const$. Thus, the zero-temperature correlation length $\xi(0)$ yields (see pages 45 and 46 in [14]):

$$\xi(0) \propto b \propto (1/\lambda_2)^{1/(2-4\hat{\lambda}_4)}. \tag{22}$$

Substituting equation (17) into equation (22) yields:

$$\xi(0) \propto T_c^{-\sigma}. \tag{23}$$

with

$$\sigma = 1/(1 - 2\hat{\lambda}_4) \tag{24}$$

being the quantum critical exponent.

By using equation (16), equation (23) can be written as the standard quantum critical scaling [23–26]:

$$\xi(0) \propto (x_o - x)^{-\nu} \tag{25}$$

with

$$\nu = m/(1 - 2\hat{\lambda}_4). \tag{26}$$

The main result of this paper is equation (23), which states that there is a scaling relationship $\xi(0) \propto T_c^{-\sigma}$ between the zero-temperature correlation length $\xi(0)$ and the transition temperature $T_c$. It is a new prediction for equation (6). To calculate the value of $\sigma$, we rewrite equation (21) in the form of a differential equation:

$$\frac{d\hat{\lambda}_4}{dlnb} \approx (3 - D) \cdot \hat{\lambda}_4 - 10\hat{\lambda}_4^2, \tag{27}$$

where we have used $z = 1$.

By equation (27), it is easy to obtain the fixed point $\hat{\lambda}_4 \approx (3 - D)/10$. Substituting it into equations (24) and (26) yields the one-loop results:

$$\sigma \approx 1 + \frac{3-D}{5}, \tag{28}$$

$$\nu \approx m\left(1 + \frac{3-D}{5}\right). \tag{29}$$

In this paper, we investigate overdoped cuprate films at $T = 0$, namely, $D = 2$. Therefore, we obtain the one-loop theoretical values:

$$\sigma \approx 1.2, \tag{30}$$

$$\nu \approx 1.2m. \tag{31}$$

Equation (23) is a novel scaling. It is derived from equation (6), where $\lambda_2(T_c)$ is quadratic in $T_c$. This differs significantly from the case of $T$-driven transitions, where [27]

$$\xi \propto T^{-\nu_0} \tag{32}$$

with

$$\nu_0 \approx \frac{1}{2} + \frac{3-D}{10}. \tag{33}$$

Furthermore, we point out that our scaling (23) differs quantitatively from the existing result. The existing literature has shown that approaching the two-dimensional-quantum-phase transition, $\xi(0)$ and $T_c$ should obey the Berezinskii–Kosterlitz–Thouless (BKT) critical behavior [25–26]:

$$\xi(0) \propto T_c^{-\frac{1}{z}}. \tag{34}$$

For $z = 1$, equation (34) implies $\sigma = 1$, which differs significantly from our one-loop result, $\sigma \approx 1.2$. To take the two-loop approximation into account, we simply observe that $\lambda_2(T_c)$ is quadratic in $T_c$ and, hence, leads to a result of 2 times Wilson's thermal critical exponent by changing $(4-D)$ with $(3-D)$. Based on this observation, equation (28) can be expanded to the two-order of $(3-D)$ [14]:

$$\sigma \approx 1 + \frac{1}{5}(3-D) + \frac{11}{100}(3-D)^2. \tag{35}$$

**Table 1.** Theoretical value for the critical exponent $\sigma$ in equation (23).

| **Critical exponent** | **The BKT result** | **Our result** |
|---|---|---|
| $\sigma$ | 1 | 1.31 |

Up to the two-loop approximation, our result (35) predicts $\sigma \approx 1.31$. Due to the difference of 0.31, it is possible to distinguish our novel result 1.31 from the BKT result 1 by experimental investigations. The comparison between our result and the BKT result is shown in Table 1. In Figure 2, we further show different variation trends of our scaling (23) and the BKT scaling (34). By contrast, our scaling (23) is expected to describe the QCP occurring in the highly overdoped side of the cuprate.

Furthermore, by equation (35) we observe that the mean-field approximation gives $\sigma = 1$, which agrees with the BKT result. Equation (12) has indicated that the mean-field approximation holds when $T_c \geq T_M$. Therefore, for two-dimensional overdoped cuprate, we should anticipate the following two-class scaling:

$$\begin{cases} \xi(0) \propto T_c^{-1}, & T_c \geq T_M \approx \frac{T_0}{1-\alpha} \\ \xi(0) \propto T_c^{-1.31}, & T_c \leq T_Q \approx \gamma(2)^2 \end{cases}.$$

[**Insert Figure 2 here**]

## 4. Experimental proposal

To test the scaling (23), we consider that a beam of $X$-rays (or neutrons) are shot perpendicularly on to a sheet of LSCO film at zero temperature ($T = 0$), as described by Figure 3. We assume that when the $X$-rays arrive in the LSCO film, they are scattered with an angle $\theta$. As with the phenomena of classical critical opalescence, by increasing the doping level $x$, we drive the transition temperature $T_c$ to approach the quantum critical point 0 so that quantum fluctuations in the LSCO film are amplified. The average length of quantum fluctuations can be denoted by the zero-temperature correlation length $\xi(0)$. This means that the LSCO film can be regarded as a 2-dimensional "plane-net" consisting of a large number of sub-cells (or lattices) with an average length $\xi(0)$. Therefore, when the $X$-rays pass through these sub-cells, the diffraction occurs, where each sub-cell can be regarded as a single slit.

According to the formula of single-slit diffraction, we should have:

$$\xi(0) \approx \frac{\lambda_X}{\theta}, \tag{36}$$

where $\lambda_X$ denotes the wavelength of the $X$-rays.

Substituting equation (36) into equation (23) yields:

$$\theta \propto T_c^\sigma. \tag{37}$$

Therefore, we can test equation (23) via investigating equation (37). By increasing the doping level $x$, one can decrease the transition temperature $T_c$. In this process, for each transition temperature $T_c^i$, one can obtain a corresponding diffraction angle $\theta^i$, where $i = 1, ..., n$. By fitting the sample data $\{T_c^i, \theta^i\}_{i=1}^n$ to equation (37), one can obtain the estimated value of the critical exponent $\sigma$.

[**Insert Figure 3 here**]

## 5. Conclusion

By introducing the imaginary time, Gor'kov's Ginzburg–Landau equation at zero temperature has been extended to an exact relativistic form (6), which is intended to describe the zero-temperature overdoped cuprate. We have shown that such a relativistic equation produces exactly the two-class scaling (1) observed in the overdoped side of single-crystal LSCO films [7]. In this paper, we further show that the relativistic equation (6) leads to a new theoretical prediction: near the superconductor–metal transition point in the overdoped side of LSCO films, the zero-temperature correlation length $\xi(0)$ and the transition temperature $T_c$ should yield a scaling $\xi(0) \propto T_c^{-\sigma}$ with a quantum critical exponent $\sigma \approx 1.31$ (up to the two-loop approximation). This differs significantly from the BKT result in the existing literature, which predicts $\sigma = 1$ for $z = 1$. To test our theoretical prediction, we propose that one can measure $\sigma$ by designing a diffraction experiment between $X$-rays and zero-temperature LSCO films, as described by Figure 3. In particular, by equation (37), we observe that the diffraction angle $\theta$ will tend to $0$ as the transition temperature $T_c$ approaches the quantum critical point $T_c = 0$. This implies a kind of phenomenon of quantum critical "opalescence."

# Appendix

By integrating out the frequency $\omega$, equations (18) and (19) can be simplified to:

$$\lambda_2'(b) = b^2 \left[ \lambda_2 + 2\lambda_4 \int_{\Lambda/b}^{\Lambda} \frac{d^D k}{(2\pi)^D} \left( \frac{1}{\sqrt{k^2+\lambda_2}} - \frac{1}{k} \right) \right], \tag{A.1}$$

$$\lambda_4'(b) = b^{4-D-z} \lambda_4 \left[ 1 - \frac{10}{4} \lambda_4 \int_{\Lambda/b}^{\Lambda} \frac{d^D k}{(2\pi)^D} (k^2 + \lambda_2)^{-\frac{3}{2}} \right], \tag{A.2}$$

where we have used the formulas

$$\int_{-\infty}^{\infty} \frac{d\omega}{2\pi} \frac{1}{\omega^2 + k^2 + \lambda_2} = \frac{1}{2\sqrt{k^2+\lambda_2}}$$

and

$$\int_{-\infty}^{\infty} \frac{d\omega}{2\pi} \left( \frac{1}{\omega^2 + k^2 + \lambda_2} \right)^2 = \frac{1}{4} (k^2 + \lambda_2)^{-\frac{3}{2}}.$$

By equation (17), near the quantum critical point $T_c = 0$, we should have $\lambda_2 \ll \Lambda$; therefore, by using the approximations $1/\sqrt{k^2 + \lambda_2} \approx (1/k) - (\lambda_2/2k^3)$ and $(k^2 + \lambda_2)^{-\frac{3}{2}} \approx 1/k^3$, equations (A.1) and (A.2) yield:

$$\lambda_2'(b) \approx b^2 \lambda_2 \left( 1 - \lambda_4 \int_{\Lambda/b}^{\Lambda} \frac{d^D k}{(2\pi)^D} \frac{1}{k^3} \right), \tag{A.3}$$

$$\lambda_4'(b) \approx b^{4-D-z} \lambda_4 \left( 1 - \frac{10}{4} \lambda_4 \int_{\Lambda/b}^{\Lambda} \frac{d^D k}{(2\pi)^D} \frac{1}{k^3} \right). \tag{A.4}$$

If we order $y = k/\Lambda$, then equations (A.3) and (A.4) can be written in the form:

$$\lambda_2'(b) \approx b^2 \lambda_2 \left( 1 - \lambda_4 \frac{S_D \Lambda^{D-3}}{(2\pi)^D} \int_{1/b}^{1} y^{D-4} dy \right), \tag{A.5}$$

$$\lambda_4'(b) \approx b^{4-D-z} \lambda_4 \left( 1 - \frac{10}{4} \lambda_4 \frac{S_D \Lambda^{D-3}}{(2\pi)^D} \int_{1/b}^{1} y^{D-4} dy \right), \tag{A.6}$$

where we have used $\int_{\Lambda/b}^{\Lambda} \frac{d^D k}{(2\pi)^D} \frac{1}{k^3} = \int_{\Lambda/b}^{\Lambda} \frac{S_D k^{D-1} dk}{(2\pi)^D} \frac{1}{k^3}$ with $S_D = 2\pi^{D/2}/\Gamma(D/2)$.

Assuming $|D - 3|$ to be sufficiently small, equations (A.5) and (A.6) yield:

$$\lambda_2'(b) \approx b^2 \lambda_2 (1 - 4\hat{\lambda}_4 \ln b), \tag{A.7}$$

$$\hat{\lambda}_4'(b) \approx b^{4-D-z} \hat{\lambda}_4 (1 - 10\hat{\lambda}_4 \ln b), \tag{A.8}$$

where $\hat{\lambda}_4 = \lambda_4 \frac{S_D \Lambda^{D-3}}{4(2\pi)^D}$.

Following the spirit of perturbation theory, we consider $\hat{\lambda}_4 \ll 1$; therefore, equations (A.7) and (A.8) can be written as:

$$\lambda_2'(b) \approx \lambda_2 b^{2-4\hat{\lambda}_4}, \tag{A.9}$$

$$\hat{\lambda}'_4(b) \approx \hat{\lambda}_4 b^{4-D-z-10\hat{\lambda}_4}. \quad (A.10)$$

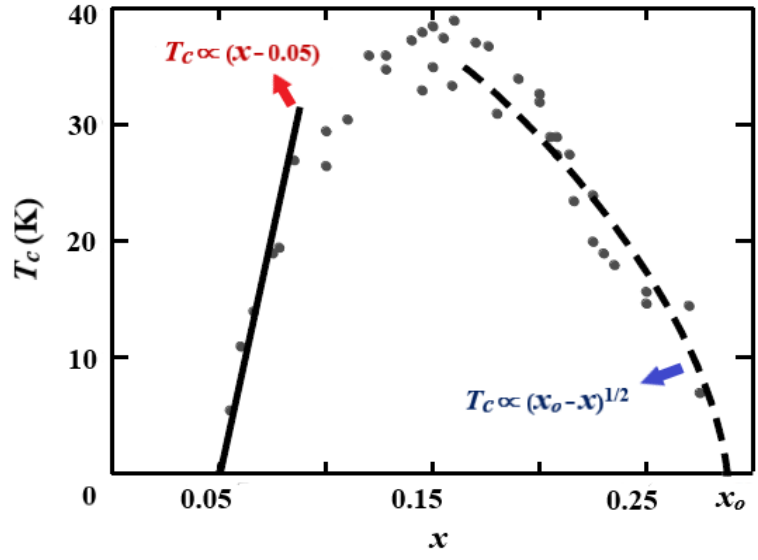

**Figure 1:** $T_c$ versus $x$ for $La_{2-x}Sr_xCuO_4$ [22]. The solid line corresponds to the asymptotic behavior of the underdoped limit with $T_c \propto (x - 0.05)$, whereas the dashed line corresponds to the asymptotic behavior of the overdoped limit with $T_c \propto (x_o - x)^{1/2}$.

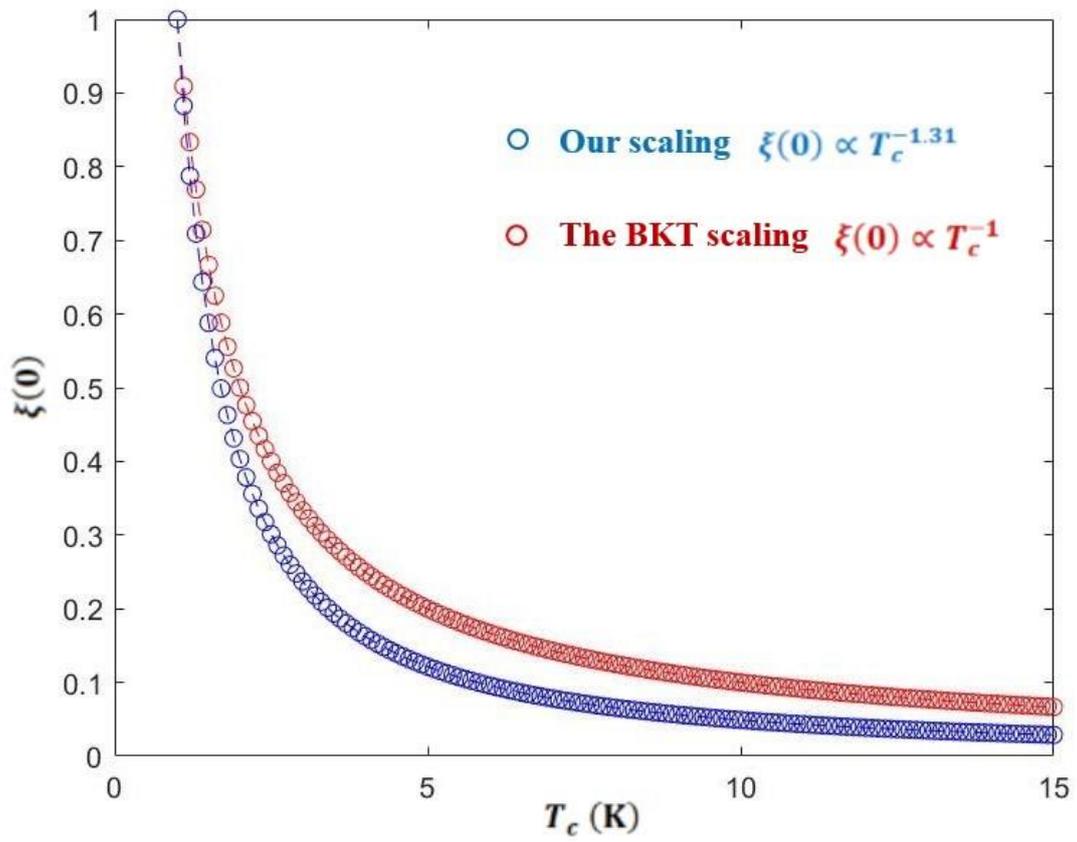

**Figure 2:** Comparison between our scaling $\xi(0) \propto T_c^{-1.31}$ and the BKT scaling $\xi(0) \propto T_c^{-1}$.

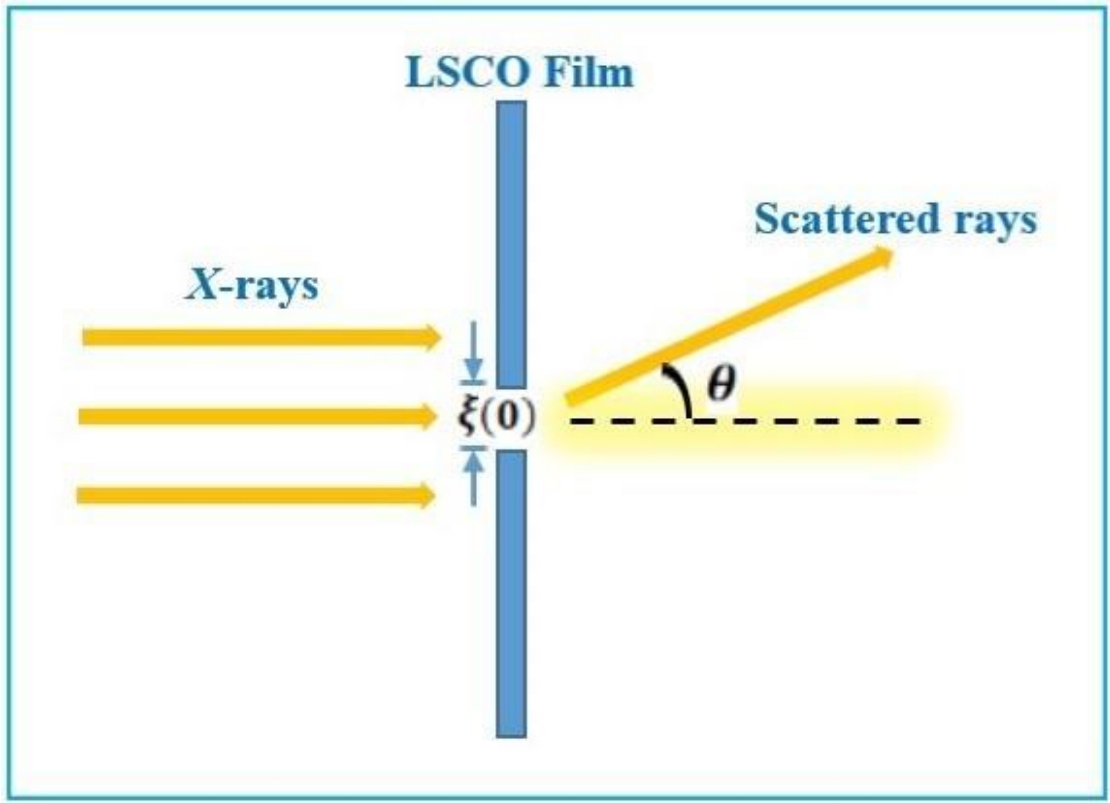

**Figure 3:** A beam of $X$-rays is shot perpendicularly on to a sheet of LSCO film at zero temperature ($T = 0$). Because $T_c = 0$ is a quantum critical point, by doping the LSCO film, one can drive the transition temperature $T_c$ to approach $0$ so that quantum fluctuations are amplified. Therefore, the zero-temperature LSCO film around $T_c = 0$ can be regarded as a 2-dimensional "plane-net" consisting of a large number of sub-cells (or lattices) with an average length $\xi(0)$, where $\xi(0)$ denotes the zero-temperature correlation length. If we regard each sub-cell as a single slit, then, by the diffraction theory of light, the $X$-rays are scattered with an angle $\theta$.